\documentclass[12pt]{article}
\usepackage{latexsym}
\usepackage{amsmath}
\usepackage{amsfonts}
\usepackage{amssymb}
\usepackage{graphicx}

\newcommand{\mR}{{\mathbb R}}

\title{Conformal Newton-Hooke algebras, Niederer's transformation and Pais-Uhlenbeck oscillator}
\author{Krzysztof  Andrzejewski
\\ \\
\small Department of  Computer Science, \\
\small University of \L\'od\'z,\\
\small Pomorska 149/153, 90-236 {\L}\'od\'z, Poland\\
\small E-mail: k-andrzejewski@uni.lodz.pl
}
\date{}
\begin{document}
\maketitle
\begin{abstract}
Dynamical systems invariant under the action of the  $l$-conformal Newton-Hooke algebras are constructed by the method of nonlinear realizations. The relevant first order Lagrangians together with  the corresponding Hamiltonians are found. The relation to the Galajinsky and Masterov  [Phys. Lett.  B 723 (2013) 190] approach as well as the higher derivatives formulation is discussed. The generalized Niederer's  transformation  are presented which relate the systems under consideration to those invariant under the action of the $l$-conformal Galilei algebra [Nucl. Phys.  B 876 (2013) 309]. As a nice application of these results an analogue of Niederer's transformation, on the Hamiltonian level,  for  the  Pais-Uhlenbeck oscillator  is   constructed.
\end{abstract}
\section{Introduction}
The Newton-Hooke algebra is a generalization  of the Galilei one to the case of nonvanishing cosmological constant  leading to the universal cosmological repulsion or attraction (see, e.g., \cite{b01,b02}).  It is derived from the (anti) de Sitter
algebra by the nonrelativistic contraction in a similar way as the Galilei algebra is obtained
from the Poincar\'e one.  The main difference between  Galilei and the Newton-Hooke algebra  is that  in the latter case the
structure relations involving the generators of time and space translations yield the Galilei
boosts: $[H, P_i] = \pm\frac{1}{R^2} K_i$. The positive constant $R$ is called the  characteristic time (and is related to the radius of the parent (anti) de Sitter space).
The upper/lower sign above is realized in nonrelativistic spacetime with the negative/positive
cosmological constant $\Lambda=\mp\frac{1}{R^2}$.
\par
Conformal extensions of the Galilei and  Newton-Hooke    algebras have recently attracted considerable interest,
mostly in the context of the nonrelativistic AdS/CFT 
correspondence. Such extensions are parameterized by a positive half-integer $l$ \cite{b03}-\cite{b06}, which
justifies  their  name: $l$-conformal algebras.  The dynamical realizations
of the $l$-conformal Galilei and Newton-Hooke  algebras involve, in general,  higher derivatives terms (see, e.g., \cite{b07b}-\cite{b19a}).
However, it is also possible  (using the method of nonlinear realizations \cite{b15a}-\cite{b15c}) to construct invariant dynamics involving only second derivatives  \cite{b23}-\cite{b27}. The method proposed  in Refs. \cite{b23}-\cite{b27} allows for elegant and algorithmic construction of invariant dynamical equations. However, there remains an open problem if they admit Lagrangian and Hamiltonian formalism.  In  Ref. \cite{b25} it has been shown that this  is possible   for the case of the $l$-conformal Galilei algebra.
\par In the present paper, first,  we apply the method developed  in  \cite{b25} to the case of the $l$-conformal Newton-Hooke algebra and  we construct the invariant dynamics in terms of the first order Lagrangian and Hamiltonian formalism (Section 2 and 3). Moreover, we compare  our approach with the one reported in Ref. \cite{b27} as well as  with the  Pais-Uhlenbeck  theory (Section 4). 
\par The second part of the paper is devoted to the problem of Niederer's-type transformations. 
In Section 5 we construct an analogon  of the celebrated Niederer's transformation \cite{b27b} for our approach, and we show that it leads to  the results in \cite{b25} obtained  for the $l$-conformal Galilei algebra. On the other hand, on the Lagrangian level, the generalization  of Niederer's   transformation 
  has been also extensively studied for the Pais-Uhlenbeck system with {\it odd} frequencies (i.e., frequencies proportional to the consecutive odd integers); see, e.g.,  \cite{b27c}-\cite{b28}. However, its Hamiltonian counterpart seems  to be more involved  due to   the lack of the direct transition to the  Hamiltonian formalism for a theory with higher derivatives.   We solve this problem and give (see, Section 6) the explicit form of the canonical transformation which relates the Pais-Uhlenbeck Hamiltonian (with odd frequencies) to the one for the free higher derivatives theory.      
 \section{Conformal mechanics}
The prototype of all conformal groups is the one acting in $1+0$-dimensional spacetime, locally isomorphic to $SL(2,\mR)$; for the recent developments in conformal mechanics see, e.g., Refs. \cite{b20a}-\cite{b20c}.
In order to construct the conformal Newton-Hooke dynamics, i.e., the dynamics of  the conformal particle in the harmonic trap, we must modify the Hamiltonian by adding the conformal generator. Thus we choose the basis of  the $sl(2,\mR)$ algebra as follows 
\begin{equation}
\label{e1}
\begin{split}
[H,D]&=i(H\mp 2K),  \\
[D,K]&=iK,  \\
[H,K]&=2iD. 
\end{split}
\end{equation}
It is worth to note that,  although we only  change  the basis ($H\rightarrow H\pm K$) of $sl(2,\mR)$ algebra, this alters the dynamics and, consequently, the dynamical realizations of the  algebra. 
\par
Let us consider the  decomposition based on $D$ as the stability subgroup generator. Then the coset space is parametrized 
as follows    
\begin{equation}
\label{e2}
\begin{split}
w=e^{itH}e^{izK},
\end{split}
\end{equation}
and the action of the $SL(2,\mR)$ group is defined by
\begin{equation}
\label{e3}
\begin{split}
ge^{itH}e^{izK}=e^{it'H}e^{iz'K}e^{iu'D},
\end{split}
\end{equation} 
which can be explicitly found by taking the  representation spaned by
\begin{equation}
H=i(-\sigma_+\pm\sigma_-),\quad K=i\sigma_-,\quad D=-\frac {i}{2}\sigma_3.
\end{equation}
It reads, 
\begin{equation}
\label{e3a}
\begin{split}
t'&=\arctan\left(\frac{\alpha \tan t+\beta}{\gamma \tan t +\delta}\right),\\
z'&=((\alpha\sin t+\beta\cos t)^2+(\gamma \sin t+\delta\cos t)^2)z,\\
& +\frac12(\beta^2+\delta^2-\alpha^2-\gamma^2)\sin 2t-(\gamma\delta+\alpha\beta)\cos 2t,\\
u'&=-\ln((\alpha\sin t+\beta\cos t)^2+(\gamma \sin t+\delta\cos t)^2),
\end{split}
\end{equation} 
in the oscillatory case ($+$),  and
\begin{equation}
\label{e3b}
\begin{split}
t'&={\textrm {arctanh}}\left(\frac{\alpha \tanh t+\beta}{\gamma \tanh t +\delta}\right),\\
z'&=((\gamma \sinh t+\delta\cosh t)^2-(\alpha\sinh t+\beta\cosh t)^2)z,\\
& +\frac12(\beta^2-\delta^2+\alpha^2-\gamma^2)\sinh 2t+(\alpha\beta-\gamma\delta)\cosh 2t,\\
u'&=-\ln((\gamma \sinh t+\delta\cosh t)^2-(\alpha\sinh t+\beta\cosh t)^2),
 \end{split}
\end{equation} 
in the hyperbolic one ($-$); here, $
g=\left(
\begin{array}{cc}
\alpha&\beta\\
\gamma&\delta
\end{array}\right)\in SL(2,\mR).
$ Due to Eq. (\ref{e1}) the  Cartan forms $w^{-1}dw\equiv i(\omega _HH+\omega _KK+\omega _DD)$ coincide with those in the old basis   and read 
\begin{align}
\label{e7}
\omega _H=dt\,\quad
\omega _K=dz+z^2dt,\quad
\omega _D=-2zdt.
\end{align}
 However, the  transformation rules change and take the form 
\begin{align}
\label{e8}
&\omega _H'=e^{u'}\omega _H,\\
&\omega _K'=e^{-u'}\omega _K+(\pm e^{-u'}\mp e^{u'})\omega_H,\\
&\omega _D'=\omega _D-du'.\nonumber
\end{align}  
The covariant derivative of $z$\ is defined as the ratio of the  Cartan forms 
\begin{equation}
\label{e9}
\begin{split}
\nabla z=\frac{\omega _K}{\omega _H}=\dot{z} +z^2 ;
\end{split}
\end{equation}  
 one can easily obtains 
\begin{equation}
\label{e10}
\begin{split}
\nabla z'=e^{-2u'}\nabla z\pm e ^{-2u'}\mp 1 .
\end{split}
\end{equation}  
 In order to construct the invariant dynamics it is sufficient to find the action integral invariant under the dilatation subgroup. This can be  easily done  by taking the Lagrangian
 \begin{equation}
\label{e11}
\begin{split}
L_0 =  {\sqrt{\dot{z}+z^2\pm 1}},
\end{split}
\end{equation} 
or the corresponding Hamiltonian
\begin{equation}
\label{e11a}
H_0=\frac{-1}{4p_z}-p_zz^2\mp p_z,
\end{equation}
with$\quad \{z,p_z\}=1$. 
The Lagrangian (\ref{e11}) (or the Hamiltonian (\ref{e11a})) leads to the following equation of motion   
 \begin{equation}
\label{e12}
\begin{split}
\ddot{z} +6z\dot{z} +4z^3\pm 4z=0 .
\end{split}
\end{equation}
The above  equation  contains the whole family of conformal models.
 In fact, with the substitution 
 \begin{equation}
\label{e13}
\begin{split}
z=\frac{\dot{\rho }}{\rho }  ,
\end{split}
\end{equation} 
suggested in Refs. \cite{b24} and \cite{b26a}, Eq. (\ref{e12}) yields
\begin{equation}
\label{e14}
\begin{split}
\frac{d}{dt}(\ddot{\rho }\rho ^3\pm\rho ^4)=0  ,
\end{split}
\end{equation}
or
\begin{equation}
\label{e15}
\begin{split}
\ddot{\rho }=\frac{\gamma ^2}{\rho ^3}\mp \rho,
\end{split}
\end{equation}
i.e., conformal particle in the harmonic trap.
\par
Let us note that the replacement
\begin{equation}
\label{e15a}
t\rightarrow  it, \quad z\rightarrow  -iz,
\end{equation}
performed in. Eq.  (\ref{e12}), relates the oscillatory case ($+$) to the hyperbolic one ($-$). This, together with the transformation rules described by the  first equation (\ref{e3a}) and (\ref{e3b}) implies the following change of the action of the $SL(2,\mR)$ group:  
\begin{equation}
\label{e15b}
g=\left(
\begin{array}{cc}
\alpha&\beta\\
\gamma&\delta
\end{array}\right) \rightarrow g'= \left(
\begin{array}{cc}
\alpha&i\beta\\
-i\gamma&\delta
\end{array}\right).
\end{equation}
Note that both realizations of  $SL(2,\mR)$ are equivalent but not related by an inner automorphism.   
\par
To get rid of the square root in the action integral one can follow the standard procedure by writing
\begin{equation}
\label{e16}
\begin{split}
L_1= -\gamma ^2\eta -\frac{1}{2\eta }(\dot{z} +z^2\pm 1)  ,
\end{split}
\end{equation}
where $\gamma$ is an arbitrary constant while $\eta $\ is an adjoint field transforming according to $\eta'=e^{-u'}\eta$.
\par
Now, let us perform a simple canonical analysis. The primary constraints read
\begin{equation}
\label{e17}
\begin{split}
\chi _1\equiv p_\eta\approx 0 , \quad \chi _2\equiv p_z+\frac{1 } {2\eta }\approx 0 ,
\end{split}
\end{equation}
 while the Hamiltonian is written as 
 \begin{equation}
\label{e18}
\begin{split}
H_1= \gamma ^2\eta +\frac{1 }{2\eta }(z^2\pm 1)+u_\eta p_\eta +u_z(p_z+\frac{1 }{2\eta }) ,
\end{split}
\end{equation} 
 $u_\eta , u_z$\ being the appropriate Lagrange multipliers. Imposing 
\begin{equation}
\label{e19}
\begin{split}
\frac{d}{dt}{p_\eta} \approx 0, \quad \frac{d}{dt}(p_z+\frac{1}{2\eta })\approx 0   ,
\end{split}
\end{equation} 
we find no new constraints while
\begin{equation}
\label{e20}
\begin{split}
u_z=2\gamma^2 \eta^2-(z^2\pm 1) , \quad u_\eta  =-2z\eta  .
\end{split}
\end{equation} 
So the constraints (\ref{e17}) are of the second kind. This allows us to eliminate $p_\eta $\ and $p_z$\ at the expense 
of introducing the Dirac bracket and finally we obtain 
\begin{equation}
\label{e22}
\begin{split}
H_D=\gamma ^2\eta +\frac{z^2\pm1}{2\eta}, \quad \{z,\eta \}_D=2\eta ^2  .
\end{split}
\end{equation} 
Putting
\begin{equation}
\label{e23}
\begin{split}
\eta =\frac{1}{\rho ^2}, \quad z=\frac{p_\rho }{\rho }  ,
\end{split}
\end{equation} 
one arrives at the standard form
\begin{equation}
\label{e24}
\begin{split}
H_D=\frac{1}{2}p_\rho ^2+\frac{\gamma ^2}{\rho ^2}\pm\frac12\rho^2, \quad \{\rho ,p_\rho \}=1  .
\end{split}
\end{equation} 
\section{Dynamical realizations of the $l$-conformal Newton-Hooke algebras}
The $l$-conformal Newton-Hooke  algebra (in three-dimensional case) is spanned
by the generators $H,D,K$ satisfying (\ref{e1})   together with $so(3)$\ generators $J_k$\ and $2l+1$ additional generators $\vec{C}^{(n)}, n=0,1,...2l$\ obeying 
\begin{align}
\label{e26}
&[H,\vec{C}^{(n)}]=i(n \vec{C}^{(n-1)}\pm(n-2l)\vec{C}^{(n+1)}),\nonumber\\
&[K,\vec{C}^{(n)}]=i(n-2l) \vec{C}^{(n+1)},\nonumber\\
&[D,\vec{C}^{(n)}]=i(n-l) \vec{C}^{(n)},\\
&[J_i,C_k^{(n)}]=i\varepsilon _{ikm} C_m^{(n)}.\nonumber
\end{align}
Consider the nonlinear action  defined by selecting the subgroup generated by $\vec{J}$\ and $D$.  
With such a choice we are not dealing with the symmetric decomposition. However, the generators $H,K$\ and $\vec{C}^{(n)}$\
span the linear representation under the adjoint action of the stability subgroup. Therefore, our realization linearizes on it. In order
 to construct the invariant dynamics it is sufficient to respect the invariance under rotations and dilatation.
 \par
 Let us choose the following parametrization of the coset manifold 
\begin{equation}
\label{e27}
\begin{split}
w=e^{itH}e^{i\vec {x}^{(n)}\vec {C}^{(n)}}e^{izK} ;
\end{split}
\end{equation}
note the difference with respect to the parametrization used in \cite{b27}.\\
The Cartan forms
\begin{equation}
\label{e28}
\begin{split}
w^{-1}dw=i(\omega _HH+\omega _DD+\omega _KK+\vec{\omega }^{(n)}\vec{C}^{(n)})  ,
\end{split}
\end{equation} 
are given by Eqs. (\ref{e7}) together with 
\begin{equation}
\label{e29}
\begin{split}
\vec{\omega }^{(n)}=\sum_{p=0}^n\dbinom{2l-p}{2l-n}(-z)^{n-p}\left( d\vec{x}^{(p)}-(p+1)\vec{x}^{(p+1)}dt\mp (p-1-2l)\vec x^{(p-1)}dt\right).
\end{split}
\end{equation}
The forms $\vec{\omega }^{(n)}$ are  vectors under $SO(3)$\ while under dilatation
\begin{equation} 
\vec{\omega }'^{(n)}=e^{(l-n)u'}\vec{\omega }^{(n)}.
\end{equation}
Define the covariant derivatives 
\begin{equation}
\label{e31}
\begin{split}
\nabla \vec{x}^{(n)}\equiv \frac{\vec{\omega }^{(n)}}{\omega _H} ,
\end{split}
\end{equation} 
with the dilatation dimension $l-n-1$. Let $\vec{\lambda }^{(n)}$\ be additional (adjoint) variables with dilatation dimension $n-l$.
Consider the following first order Lagrangian
\begin{equation}
\label{e32}
\begin{split}
L=-\gamma ^2\eta -\frac{1}{2\eta }(\dot{z}+z^2\pm1) +\sum_{n=0}^{2l} \vec{\lambda }^{(n)}\nabla \vec{x}^{(n)}.
\end{split}
\end{equation} 
By the very construction it yields the invariant action functional. The equations of motion are of the form
\begin{align}
\label{e33}
&2\gamma ^2\eta ^2-(\dot{z}+z^2\pm 1)=0,\nonumber \\
&\dot{\eta }+2z\eta =0,\nonumber \\
&\dot{\vec{x}}^{(n)}-(n+1)\vec{x}^{(n+1)}\mp\vec{x}^{(n-1)}(n-1-2l)=0, \quad n=0,...,2l , \\
&\sum_{n=0}^{2l-p}\dbinom{2l-p}{n}\frac{d}{dt}\left((-z)^n\vec{\lambda }^{(n+p)}\right)+p\sum_{n=0}^{2l-p+1}
\dbinom{2l-p+1}{n}(-z)^n\vec{\lambda }^{(n+p-1)}\nonumber\\
&\pm (p-2l)\sum_{n=0}^{2l-p-1}
\dbinom{2l-p-1}{n}(-z)^n\vec{\lambda }^{(n+p+1)}=0. \nonumber
\end{align} 
We see that they  decouple.   The first two describe the conformal mechanics in the harmonic trap. Then, there is a set of equations for
$\vec{x}^{(n)}$\ describing  higher derivatives system. Let us note that in our approach   we do not need to perform the redefinition of time 
as in  Ref. \cite{b27}. Finally, once $z(t)$\ is determined one can solve the last equation
for $\vec{\lambda }^{(n)}$; they do not impose any further constraints on $z$.  
\par
Finally, let  us note that extending the transformation rules (\ref{e15a})   by 
\begin{equation}
\eta \rightarrow -i\eta,\quad \vec \lambda_p\rightarrow i^p\vec\lambda_p,\quad \vec x_p\rightarrow (-i)^p\vec x_p,
\end{equation} 
one can transform the Lagrangian (\ref{e32}) from the oscillatory to the  hyperbolic case. 
\par
Our Lagrangian, being of the first order, provides an example of a constrained system. Following \cite{b27a},  the Hamiltonian dynamics is given by
\begin{equation}
\label{e45}
H=\gamma ^2\eta +\frac{z^2\pm1}{2\eta } + \sum_{n=0}^{2l}\vec{\lambda }^{(n)}\sum_{p=0}^n\dbinom{2l-p}{2l-n}(-z)^{n-p}\Big((p+1)\vec{x}^{(p+1)}
\pm(p-1-2l)
\vec{x}^{p-1}\Big),
\end{equation}
together with
\begin{equation}
\label{e46}
\begin{split}
\{x_a^{(n)}, \lambda ^{(m)}_b\}_D&=z^{n-m}\dbinom{2l-m}{2l-n}\delta _{ab} ,\\
\{z,\eta \}_D&=2\eta ^2, \\
\{\vec{\lambda }^{(k)},\eta \}_D&=2(2l-k)\eta ^2\vec{\lambda }^{(k+1)} .
\end{split}
\end{equation}
Again, it is straightforward, although slightly tedious, to check that Eqs.  (\ref{e45}) and (\ref{e46}) yield the correct dynamics.
\section{Pais-Uhlenbeck oscillator} 
 It has been shown in Ref.  \cite{b28}   that for $l$  half-integer  (i.e. $2l$ is odd) the Pais-Uhlenbeck oscillator  of order $2l+1$ \cite{b29}
\begin{equation}
\label{e51b}
L=\frac{(-1)^{l+\frac 12}}{2}\vec x \prod_{k=1}^{l+\frac{1}{2}}\left(\frac{d^2}{dt^2}+\omega_k^2\right)\vec x, 
\end{equation}
with odd frequencies  $\omega_k=(2k-1)\omega=(2k-1)$ ({in what follows we put $\omega=\frac{1}{R}=1$}), $k=1,2,\ldots,l+\frac{1}{2}$  
  enjoys $l$-conformal Newton-Hooke symmetry (in fact,  it is the maximal  symmetry group). 
\par
In order to compare this finding with our results, let us  note that for the $l$ half-integer,  we  can put the oscillator system defined by the decoupled equations for  $\vec x$'s 
\begin{equation}
\label{e51a}
\dot{\vec{x}}^{(n)}-(n+1)\vec{x}^{(n+1)}\mp\vec{x}^{(n-1)}(n-1-2l)=0, \quad n=0,...,2l , \\
\end{equation}
 into the unconstrained Hamiltonian form. To see this we define 
 \begin{align}
\label{e54}
H&=\sum_{k=0}^{l-\frac{3}{2}}\vec{p}_k\vec{q}_{k+1}+\frac{1}{2}\vec{p}_{l-\frac{1}{2}}^{\;2}
\pm\left(-\sum_{k=0}^{l-\frac{3}{2}}(k+1)(2l-k)\vec{q}_{k}\vec{p}_{k+1} \right.\nonumber \\
 &+\left.\frac{1}{2}(l+\frac{1}{2})^2\vec{q}_{l-\frac{1}{2}}^{\;2}\right),
\end{align}
which corresponds to our change of the basis  $H\rightarrow H\pm K$ in the algebra of the free theory;    the standard Poisson brackets read
\begin{equation}
\label{e52}
\{q_{ka},p_{jb}\}=\delta _{kj}\delta _{ab}.
\end{equation}
The Hamiltonian (\ref{e54}) together with the Poisson brackets (\ref{e52}) yield  the  following equations of motion 
\begin{equation}
\begin{split}
&\dot{\vec q}_k=\vec q_{k+1}\mp(2l+1-k)k\vec q_{k-1},\\
&\dot {\vec p}_k=\pm(k+1)(2l-k)\vec p_{k+1}-\vec p_{k-1},\\
&\dot {\vec q}_{l-\frac12}=\vec p_{l-\frac12}\mp(l+\frac32)(l-\frac12)\vec q_{l-\frac32},\\
&\dot {\vec p}_{l-\frac12}=\mp{(l+\frac12)}^2\vec q_{l-\frac 12}-\vec p_{l-\frac32},
\end{split}
\end{equation}
for $ k=0,....l-\frac{3}{2}$; which after making the substitution
\begin{equation}
\label{e51}
\begin{split}
&\vec{q}_k=k!\vec{x}^{(k)}, \\
&\vec{p}_k=(-1)^{l-\frac{1}{2}-k}(2l-k)!\vec{x}^{(2l-k)},
\end{split}
\end{equation} 
for  $ k=0,....l-\frac{1}{2} $,  become equivalent to Eqs. (\ref{e51a}). 
 \par
On the other hand, let us observe  (see, \cite{b30} and \cite{b31}) that the Hamiltonian (\ref{e54})  is related (in the ($+$) case) through a canonical transformation to the one  for the Pais-Uhlenbeck Lagrangian (\ref{e51b}), i.e.,  
\begin{equation}
\label{e51c}
H=\sum_{k=1}^{l+\frac 12}\frac{(-1)^{l+\frac 12-k}}{2}\left({\vec  P_k}^2+(2k-1)^2{\vec Q_k}^2\right).
\end{equation}
So, in the case of $l$\,half-integer there exists an alternative Hamiltonian formalism with no additional variables. On the  contrary, for  $l$ integer   the auxiliary dynamical variables  $\vec\lambda$'s are necessary.
\section{Generalized  Niederer's  transformation }
As it was mentioned before, the $l$-conformal Newton-Hooke algebra is a counterpart  of the $l$-conformal Galilei one in the presence
of a universal cosmological repulsion or attraction. Since  these algebras are isomorphic we expect that they dynamical realizations should be related to each other  in analogy   to  the case of $l=\frac12$, where their realizations (motion of the free particle and a half-period motion of the harmonic oscillator) are  related by famous Niederer's transformation  \cite{b27b}.  In this section we will  show  that the realizations obtained in the  preceding  sections  are also related, by a counterpart of Niederer's transformation, to the ones obtained in \cite{b25} for  the $l$-conformal Galilei algebra.
\par It is worth to notice that this fact  holds  for both $l$ integer and half-integer. However, in the second case (as we saw in the preceding section) we have at our disposal an alternative Hamiltonian formalism -- the Pais-Uhlenbeck Hamiltonian with odd frequencies.  In the next section, we will apply the results obtained here to that important case.
\par     
First, let us denote with tilde  the dynamical variables entering the  realizations of the $l$-conformal Galilei  algebra\footnote{However, for simplicity, the derivatives with respect to $\tilde t$ are also denoted by dots.} and  define
\begin{equation}
\label{e46a}
\tilde\kappa  (\tilde t)=\left\{
\begin{array}{c}
\sqrt{1+\tilde t^2}\qquad  (+) \textrm{ oscillatory case},\\
\sqrt{1-\tilde t^2} \qquad  (-) \textrm{ hyperbolic case}.
\end{array}\right.
\end{equation}
Then $\tilde\kappa$ satisfies the following useful relations
\begin{equation}
\label{e46b}
\dot{\tilde\kappa}\tilde\kappa=\pm\tilde t, \quad \dot{\tilde\kappa}^2=\pm1\mp\frac{1}{\tilde\kappa^2}, \quad \overset{..}{\tilde\kappa}=\pm\frac{1}{\tilde\kappa}-\frac{\dot{\tilde\kappa}^2}{\tilde\kappa} ,
\end{equation}
and, consequently, the  equation of motion for the conformal mechanics $   \overset{..}{\tilde\kappa}=\pm\frac{1}{\tilde\kappa^3}$.
\par
Now, we define  a counterpart of  Niederer's transformations as follows 
\begin{equation}
\label{e46c}
\begin{split}
&t=\arctan \tilde t,  \quad (+) \textrm { case}; \quad  t=\textrm{arctanh}\, \tilde t,  \quad (-) \textrm { case};\\
&z=\tilde\kappa^2\tilde z -\dot{\tilde\kappa}\tilde\kappa.
\end{split}
\end{equation}
First, by the direct calculations, we can check that the  action of the  $SL(2,R)$ group  on $(t,z)$ (Eqs.  (\ref{e3a}) and (\ref{e3b})) transforms into the one for $(\tilde t, \tilde z)$, (cf. Ref.  \cite{b25}).
Next, we verify that   the Lagrangian (\ref{e11}) transforms exactly (no total time derivative is needed)  into the one obtained in \cite{b25}, i.e.,
\begin{equation}
\label{e46d}
\tilde L_0=\sqrt{\dot {\tilde {z}}+\tilde z^2}.
\end{equation}
The same situation occurs on the Hamiltonian level. Indeed, defining 
\begin{equation}
\label{e46dd}
p_z=\frac{\tilde p_z}{\tilde\kappa^2},
\end{equation} 
we obtain {the  time dependent} canonical transformation, which transforms  the Hamiltonian (\ref{e11a}) into the  conformal one, i.e.,
\begin{equation}
H_0\frac{d  t}{d \tilde t}+\frac{\partial F}{\partial t}=\frac{-1}{4\tilde p_z}-\tilde p_z\tilde z^2=\tilde H_0,
\end{equation}
 where 
\begin{equation}
 \label{e46ddd}
 \frac{d  t}{d \tilde t}=\frac{1}{\tilde\kappa^2},
 \end{equation}
while $F_0(z,\tilde p_z,\tilde t)=\tilde p_z(z\tilde\kappa^{-2}+\dot{\tilde\kappa}\tilde\kappa^{-1})$ is the  generating function for  the transformation (\ref{e46c}) and  (\ref{e46dd}).
 
\par
Moreover, adding the following transformation rule for the dynamical variable $\eta$
\begin{equation}
\label{e46e}
\eta=\tilde\kappa^2\tilde\eta,
\end{equation}
 we obtain the generalization of Niederer's transformation for the Lagrangian (\ref{e16}).	
\par 
Next,  we supply the  transformations (\ref{e46c}) and (\ref{e46e}) by the ones  for the remaining  dynamical variables 
\begin{equation}
\label{e46f}
\begin{split}
\vec x^{(p)}&=\sum_{m=0}^p\dbinom{2l-m}{2l-p}(-\dot {\tilde\kappa})^{p-m}\tilde\kappa^{m+p-2l}\tilde{\vec  x}^{(m)},\\
\vec \lambda^{(p)}&=\tilde\kappa^{2l-2p}\tilde{\vec\lambda}^{(p)},
\end{split}  
\end{equation}
where  $ p=0,\ldots,2l.$ 
Now, making the substitution  defined by Eqs.  (\ref{e46c}), (\ref{e46e}) and (\ref{e46f}) in  the Lagrangian (\ref{e32}) and  using Eqs. (\ref{e46b}) together   with the following identities
\begin{equation}
\label{e46g}
\begin{split}
&0=(m-p)\dbinom{2l-m}{2l-p}+(2l-p+1)\dbinom{2l-m}{2l-p+1},\\
&0=m\dbinom{2l-m+1}{2l-p}-(p+1)\dbinom{2l-m}{2l-p-1}+(2l-m-p)\dbinom{2l-m}{2l-p},
\end{split}
\end{equation}
we arrive,  after straightforward but rather tedious computations, at the Lagrangian invariant under  the action of the  $l$-conformal Galilei algebra (see, \cite{b25})
 \begin{equation}
\label{e46h}
\begin{split}
\tilde L=-\gamma ^2\tilde \eta -\frac{1}{2\tilde \eta }(\dot{\tilde z}+\tilde z^2) +\sum_{n=0}^{2l}\sum_{p=0}^n \tilde{\vec{\lambda }}^{(n)}
\dbinom{2l-p}{2l-n}(-\tilde z)^{n-p}\left( \dot{\tilde{\vec{x}}}^{(p)}-(p+1)\tilde{\vec{x}}^{(p+1)}\right).
\end{split}
\end{equation} 
Let us stress that there is no total time derivative entering the transformation rule.    
\section{Niederer's transformation for Pais-Uhlenbeck model 		on the Hamiltonian level}
Let us recall  (see, Ref. \cite{b28}) that the Pais-Uhlenbeck oscillator   described by the Lagrangian (\ref{e51b}) is related  to the  free higher derivatives theory, defined by the Lagrangian
\begin{equation}
\label{e53}
\tilde L=\frac 12 \left( \frac{d^{l+\frac 12}\tilde{\vec{x}}}{d^{l+\frac 12}\tilde t} \right)^2.
\end{equation}
The relevant transformation  reads    
\begin{equation}
t=\arctan \tilde t, \quad \vec x={\tilde\kappa^{-2l}}{\tilde{\vec x }}  .
\end{equation}
However,  passing to the Hamiltonian counterpart of this transformation  we encounter some difficulties; there is no straightforward  transition to the Hamiltonian formalism  for Lagrangians with higher derivatives  (in general, we have to introduce some   auxiliary     variables and next apply the Dirac's method for constraint systems). 
We will fill this gap below. Namely, using  the results from the preceding sections, we construct  a canonical transformation relating the Hamiltonian (\ref{e54}) to the one corresponding to the  free  theory, i.e., the Ostrogradski Hamiltonian corresponding to  the Lagrangian (\ref{e53}):
\begin{align}
\label{e55}
\tilde H&=\sum_{k=0}^{l-\frac{3}{2}}\tilde{\vec{p}}_k\tilde{\vec{q}}_{k+1}+\frac{1}{2}\tilde{\vec{p}}_{l-\frac{1}{2}}^{\;2}.
\end{align}     
\par
We will work in terms of  the variables $q$'s and $p's$ and Hamiltonian (\ref{e54}) since in this approach the Pais-Uhlenbeck Hamiltonian (for odd frequencies)  is the sum of the  Hamiltonian  and the conformal generator (at time zero) of the free theory which perfectly  corresponds with the relation between the $l$-conformal Galilei  and Newton-Hook algebra.  An explicit form of the canonical transformation  between   $q$'s and $p$'s   and the decouple  harmonic variables as well as Ostrogradski  ones will be given  in the forthcoming paper \cite{b31}; what enables to find  this transformation in both  remaining approaches.  
\par        
 Let us start with the  crucial observation  that the relations (\ref{e51}) can be used also in the case of the  free theory and  that    Eqs. (\ref{e46f})  define  Niederer's-type  transformation in our Lagrangian formalism (with no   total time derivative entering). Following this idea we obtain the transformation 
  \begin{equation}
  \label{e56}
  \begin{split}
  \vec q_k&=\sum_{m=0}^{l-\frac 12}b_{km}\tilde{\vec {q}}_m ,\\
  \vec p_k&=\sum_{m=0}^{l-\frac l2}(b^{-1})_{mk}\tilde{\vec p}_m+\sum_{m=0}^{l-\frac l2}c_{mk}\tilde{\vec  q}_m,
  \end{split}
  \end{equation}
  where 
   \begin{equation}
  \label{e57}
  \begin{split}
  b_{km}&=\frac{k!}{m!}\dbinom{2l-m}{2l-k}(-\dot{\tilde\kappa})^{k-m}\tilde\kappa^{m+k-2l},\\
  c_{mk}&=\frac{(2l-k)!}{m!}(-1)^{ l-\frac12-k}\dbinom{2l-m}{k}(-\dot{\tilde\kappa})^{2l-k-m}\tilde\kappa^{m-k}, \\
    (b^{-1})_{mk}&=(-1)^{k+m}\tilde\kappa^{4l-2m-2k}b_{mk}
  \end{split}
  \end{equation}
and, by definition, $\dbinom{k}{m}=0$ if $k<m$. We will check that Eqs.  (\ref{e56}) define, on the Hamiltonian level,  an analogue (to the classical case $l=\frac 12$) of  Niederer's   transformation  relating Pais-Uhlenbeck model with odd frequencies and the free higher derivatives theory, i.e.,
\begin{equation}
\label{e58}
H\frac{d  t}{d \tilde t}+\left(\frac{\partial F}{\partial \tilde t}\right)=\tilde H ;
\end{equation}
where $F$ is the generating function for the transformation (\ref{e56}) and both sides are expressed in terms of $\tilde q$'s and $\tilde p$'s\footnote{Eq. (\ref{e58}) is the well know transformation rule for the  Hamiltonian, under a canonical transformation,  in the case when time variable.}.
\par 
First,  by the standard calculations  we check that Eqs. (\ref{e56}) define a canonical transformation.
Further, we find  the generating function 
\begin{equation}
\label{e60}
F(\vec q_0,\ldots,\vec q_{l-\frac12},\tilde{\vec p}_0,\ldots,\tilde{\vec p}_{l-\frac 12},\tilde t)=\sum_{k=0}^{l-\frac12}\tilde{\vec p}_k\tilde{\vec q}_k(\vec q_0,\ldots,\vec q_{l-\frac12},\tilde t)+\frac 12 \sum_{k,m=0}^{l-\frac12}a_{km}\vec q_k\vec q_m,
\end{equation}
where
\begin{equation}
\label{e61}
a_{km}=\frac{(-1)^{k+m}(2l-k)!(2l-m)!}{k!m!(l-\frac 12 -k)!(l-\frac12-m)!}\frac{(		-\tilde\kappa\dot{\tilde\kappa})^{2l-k-m}}{(2l-k-m)}=a_{mk},
\end{equation} 
 and, by virtue of  (\ref{e56})
\begin{equation}
\tilde{\vec q}_m(\vec q_0,\ldots,\vec q_{l-\frac12},\tilde t)=\sum_{k=0}^{l-\frac12}(b^{-1}\left(\tilde t)\right)_{mk}\vec q_k.
\end{equation}
 To this end the identity 
\begin{equation}
\label{e59}
\sum_{k=0}^a(-1)^k\dbinom{a+b}{k}=(-1)^a\dbinom{a+b-1}{a}, \quad 0\leq a,\, 1\leq b ,
\end{equation}
appears to be very useful.
Next, we prove Eq. $(\ref{e58}) $: due to the fact  that the Pais-Uhlenbeck model is traditionally  considered in the oscillatory regime (cf. Eqs. (\ref{e51b}) and (\ref{e51c})),  we will focus on the ($+$) case; the $(-)$ case can be treated in the same way or by using the observation that 
the transformation
\begin{equation}
\label{e62}
\vec q_k\rightarrow (-i)^{l+\frac12 +k}\vec q_k ,\quad \vec p_k\rightarrow (-1)^{l+\frac 12}(-i)^{l -\frac 12 -k}\vec p_k ,
\end{equation}   
relates   ($+$) and ($-$) cases.
\par   	 
Using Eq. (\ref{e46b})   as well as the  known properties of the binomial coefficients we obtain, after straightforward  but rather tedious computations, the derivative of $F$  with respect to $\tilde t$ --  expressed in terms of $\tilde q$'s  and $\tilde p$'s:
\begin{equation}
\label{e63}
\begin{split}
&\frac {\partial F}{\partial \tilde t} (\tilde{\vec q}_0,\ldots,\tilde{\vec q}_{l-\frac12},\tilde{\vec p}_0,\ldots,\tilde{\vec p}_{l-\frac 12},\tilde t)=\frac{1}{\tilde\kappa^ 2}\sum_{m=0}^{l-\frac 32}(2l-m)(m+1) \tilde{\vec p}_{m+1}\tilde{\vec q}_m\\
&+\frac{2\dot {\tilde\kappa}}{\tilde\kappa}\sum_{m=0}^{l-\frac 12}(l-m)\tilde{\vec p}_m\tilde{\vec q}_m-\frac{1}{2\tilde\kappa^2}(l+\frac 12)^2{\tilde {\vec q}}_{l-\frac 12}^2.
\end{split}
\end{equation}
So, to prove Eq.  (\ref{e58}) it remains to express $H$ in terms of $\tilde q$'s and  $\tilde p$'s. The explicit  calculations are 	troublesome so we will sketch only  the main steps.
\par  First we find the  coefficients in front of  the terms  $\tilde{\vec  q}_m\tilde{\vec  q}_{\bar m}$. 
Using Eq. (\ref{e46b}) and the identities  (\ref{e46g}) we derive the following relations
\begin{equation}
\label{e64}
(k+1)(2l-k)c_{m,k+1}-c_{m,k-1}=\tilde\kappa^2c_{m-1,k}-2\tilde\kappa\dot{\tilde\kappa}(l-m)c_{mk}-(2l-m)(m+1)c_{m+1,k},
\end{equation}
for $k,m=0,\ldots,l-\frac12$.
Next, applying the identity (\ref{e59}),  we compute the expressions of the type  $\sum_{k=m}^{l-\frac 12}c_{\bar m k}b_{km}$. Due to the  symmetry $m\leftrightarrow \bar m$ the final result is of the form $ \frac{1}{2}(l+\frac 12)^2{\tilde {\vec q}}_{l-\frac 12}^2$, and by Eq. (\ref{e46ddd}) it cancels against the last term of Eq. (\ref{e63}).
\par To compute the  coefficients in front of the  terms bilinear in  $\tilde q$'s and $\tilde p's$, we first derive, by virtue of Eqs. (\ref{e46b})  and   (\ref{e46g}) the following identity  
 \begin{equation}
 \label{e65}
b_{k+1,m}-k(2l-k+1)b_{k-1,m}=\tilde\kappa^2b_{k,m-1}-2\tilde\kappa\dot{\tilde\kappa}(l-m)b_{km}-(2l-m)(m+1)b_{k,m+1},
\end{equation}
for $m=0,\ldots,l-\frac 12$ and $k=0\,\ldots,l+\frac 12$.  Using  (\ref{e65}) and (\ref{e46ddd}) we conclude that the final result contains  three  terms:  two of them  cancel against the  first two terms of    Eq. (\ref{e63}) and there only remains the sum  $\sum_{m=0}^{l-\frac{3}{2}}\tilde{\vec{p}}_m\tilde{\vec{q}}_{m+1}$.  Finally, it is quite easy to check that the only nonvanishing term bilinear in $\tilde p'$s  is $\frac{1}{2}\tilde{\vec{p}}_{l-\frac{1}{2}}^{\;2}$. In summary, we obtain the Hamiltonian $\tilde H$   (cf. Eq. (\ref{e55}))  and, consequently, the relation (\ref{e58}).
\section{Conclusions} 
\par
 We have used the method of the nonlinear realizations to construct dynamical systems invariant under the action of the $l$-conformal Newton-Hooke algebra for  both integer and half-integer values of $l$. We put emphasis on the  Lagrangian and Hamiltonian formulation. Therefore, instead of imposing invariant constraints on the Cartan forms we enlarged the stability subgroup  (in order to abandon one constraint) and added new variables which allow us to construct  a simple invariant Lagrangian in such a way that these new degrees of freedom do not enter the dynamics of the  original ones. The resulting dynamical equations of motion are described by Eqs. (\ref{e33}).  The characteristic property of  Eqs. (\ref{e33})  is that they  decouple. We have achieved this by the appropriate choice of the subgroup, on which the action of the $l$-conformal group linearizes (rotations and dilatation) and the specific parametrization of the coset manifold (cf. Eq. (\ref{e27})).
\par We have shown that this description is universal in the sense that it works whether $l$ is half-integer or integer. The difference between the case of $l$ integer or half-integer is that the latter admits, besides the  Hamiltonian formalism presented here, an alternative one where no additional variables are necessary, namely, the Hamiltonian formalism of the  Pais-Uhlenbeck oscillator with odd frequencies. Note that, when $\vec \lambda^{(n)}$ variables  are present, the group action is no longer transitive and the phase space is not a coadjoint orbit and cannot be directly obtained by the orbit method. 
\par
Next, we constructed an analogy of  Niederer's transformation relating the dynamics described in Section 2 and 3  to the one constructed in Ref. \cite{b25} for the  $l$-conformal Galilei algebra. Moreover, we use this transformation as well as the relations between our  Lagrangian formalism  and the Pais-Uhlenbeck theory to find the counterpart of   Niederer's  transformation for the  Pais-Uhlenbeck oscillator on the Hamiltonian level.   This is accomplished by the canonical transformation (\ref{e56}). We believe that  this transformation can be useful  to extend  Niederer's transformation to the quantum  version of the   Pais-Uhlenbeck model  as well as the study of its quantum symmetries.  It is also  tempting (especially in the context of the recent results \cite{b32}) 
to extend the present considerations to the supersymmetric case: in particular, to find supersymmetric extensions of Niederer's transformations.
\vspace{0.5cm}
\par
{\bf Acknowledgments.}
Special thanks are to Piotr Kosi\'nski for valuable comments and suggestions. 
The discussions  with Joanna Gonera and Pawe\l\  Ma\'slanka are  gratefully acknowledged.
The work is supported by the grant of National Research Center number 
DEC-2013/09/B/ST2/02205.

\end{document}